\documentclass{aa}
\usepackage{graphicx}
\usepackage{txfonts}
\usepackage{xcolor}
\usepackage[switch]{lineno}
\begin{document}

\title{An Algorithm to locate the centers of Baryon Acoustic Oscillations}

\author{Z. Brown
\and G. Mishtaku
\and R. Demina
\and Y. Liu
\and C. Popik}

\institute{Department of Physics and Astronomy, University of Rochester,\\
500 Joseph C. Wilson Boulevard, Rochester, NY 14627, USA\\
\email{zbrown5@ur.rochester.edu}}

\date{Received XX XX, XXXX; accepted YY YY, YYYY}

\abstract{The cosmic structure formed from Baryon Acoustic Oscillations (BAO) in the early universe is imprinted in the galaxy distribution observable in large scale surveys, and is used as a standard ruler in contemporary cosmology. BAO are typically detected as a preferential length scale in two point statistics, which gives little information about the location of the BAO structures in real space.}{The aim of the algorithm described in this paper is to find probable centers of BAO in the cosmic matter distribution.}{The algorithm convolves the three dimensional distribution of matter density with a spherical shell kernel of variable radius placed at different locations. The locations that correspond to the highest values of the convolution correspond to the probable centers of BAO. This method is realized in an open-source, computationally efficient algorithm.}{We describe the algorithm and present the results of applying it to the SDSS DR9 CMASS survey and associated mock catalogs.}{A detailed performance study demonstrates the algorithm's ability to locate BAO centers, and in doing so presents a novel detection of the BAO scale in galaxy surveys.}

\keywords{cosmology: observations -- large-scale structure of Universe --  dark energy -- dark matter}

\maketitle

\section{Introduction}
\label{intro}

Baryon acoustic oscillations (BAO) are density waves which formed in the photon-baryon plasma in the primordial universe \citep{sunyaev1970small,peebles1973statistical,eisenstein1998baryonic,bassett2009baryon}. They are the result of the competition between the gravitational attraction pulling matter (mostly dark matter) into regions of high local density and radiation pressure pushing baryonic matter away from regions of high density. The resulting density waves, propagating at the sound speed of the plasma, produced ‘bubbles’ with high-density centers, relatively underdense interiors, and overdense spherical ’shells.’ At the time of recombination when photons and matter fell out of thermal equilibrium, the bubbles ‘froze’ into the matter distribution, with the centers remaining enriched in dark matter and shells enriched in baryonic matter \citep{eisenstein2007robustness,tansella2018second}. After recombination the BAO centers and shells became the seeds of galaxy formation.

At late times the BAO can be observed as a preferred length scale in the distribution of galaxies using the two-point correlation function (2pcf) and its Fourier transform, the power spectrum \citep{eisenstein2005detection,percival2007measuring}, as well as the three-point correlation function (3pcf) and its Fourier transform, the bispectrum \citep{slepian2017large,slepian2017detection}. Since the BAO signal is a small perturbation in the large-scale clustering of galaxies, it is observed on a statistical basis. The positions of individual BAO centers and shells are not typically identified in large galaxy surveys. However, the identification of BAO centers would be of significant cosmological and astrophysical interest. For example, the cross-correlations between BAO centers and other tracers of large-scale structure, such as galaxies, voids, and dark matter tracers found via weak lensing, could probe different models of structure formation. Moreover, the identification of individual BAO centers and shells can address questions not currently answerable with the two and three-point statistics, such as the average number of galaxies in the shells.

The construction of higher-order correlation functions is computationally expensive, so we propose a method called {\tt CenterFinder}\footnote{The source code may be downloaded from \url{https://github.com/mishtak00/centerfinder}} to locate the centers of BAO bubbles and number of galaxies $N$ displaced from the center by a distance $R_\mathrm{BAO}$. The {\tt CenterFinder} algorithm is inspired by a template track-finding algorithm originally suggested in \cite{hough1962} and generalized in \cite{ballard1981}, which is typically used in particle physics (see e.g., \cite{demina2004measurement}). In the following section, we describe the design and free parameters of the {\tt CenterFinder} algorithm. In Section~\ref{sec:performance} we study the performance of {\tt CenterFinder} using mock galaxy catalogs and the SDSS DR9 CMASS galaxy survey. We discuss the results in Section~\ref{subsec:discussion} and conclude in Section~\ref{sec:conclusion}.

\section{The CenterFinder Algorithm}
\label{sec:description}

{\tt CenterFinder} locates BAO clustering centers by convolving 3D spherical kernels of adjustable radii with tracers of large-scale structure. In this section we describe the inputs to the algorithm (Section \ref{subsec:inputcoords}), the estimator of the local density field (Section \ref{subsec:density}), the definition of the kernel (Section \ref{subsec:kernel}), the convolution step (Section \ref{subsec:conv}), and the output (Section \ref{subsec:output}). 

\subsection{Input and Coordinates}
\label{subsec:inputcoords}

{\tt CenterFinder} takes as its input catalogs of various matter tracers from surveys or simulations. For each tracer, its right ascension, $\alpha_g$, declination, $\delta_g$ and redshift, $z_g$ are required. Tracer weights, $w_g$ may be used, but are not essential. Within the algorithm, celestial coordinates are converted to 3D Cartesian coordinates. The relationship between the redshift and the comoving radial distance, $r_g$ depends on the assumed cosmology:

\begin{align}
r_g(z_g) = \frac{c}{H_0} { \displaystyle\int_{0}^{z_g} } \frac{dz'}{ \sqrt{\Omega_M(z'+1)^3+\Omega_k(z'+1)^2+\Omega_\Lambda} }\ , 
\label{eq:hubbleint}
\end{align} 
where $\Omega_M$, $\Omega_k$, and $\Omega_\Lambda$ are the present-day values of the relative densities of dark matter, spatial curvature, and dark energy, respectively. $H_0$ is the present day Hubble's constant and $c$ is the speed of light. The integral in Eq.~\ref{eq:hubbleint} is evaluated numerically in {\tt CenterFinder}. The Cartesian coordinates of each tracer are then evaluated according to:

\begin{align}
X_g=r_g \cos(\delta_g)\cos(\alpha_g)\ , 
\label{eq:cartesianx} \\
Y_g=r_g \cos(\delta_g)\sin(\alpha_g)\ ,
\label{eq:cartesiany} \\
Z_g=r_g \sin(\delta_g)\ .
\label{eq:cartesianz}
\end{align}
To prevent confusion with the redshift, $z$, Cartesian coordinates are given as capital letters, $X$, $Y$, $Z$.

\subsection{The Density Field} 
\label{subsec:density}

Several methods of estimating the density field may be used in this algorithm. The details are provided in the associated {\tt README} document. These methods rely on either raw or weighted galaxy histograms, as well as the galaxy over-density with respect to the survey mean density. Here we describe the option, which is realized in this study. We start by defining a grid with spacing $d_c$, such that the volume of each grid cell is $d_c^3$. $X_i,Y_j,Z_k$ denote the Cartesian coordinates of the $i, j, k$-th cell. On this grid we define 3-dimensional histograms $N_{wtd}(X_i,Y_j,Z_k)$, which denotes a number count of tracers, and $R(X_i,Y_j,Z_k)$, which represents a number count of randomly distributed points within the same fiducial volume: 
\begin{align}
R(X_i,Y_j,Z_k) = R(\alpha_i,\delta_j,z_k) \frac{dXdYdZ}{dV_{sph}} \ , 
\label{eq:randcartesian}
\end{align} 
where $dXdYdZ$ and $dV_{sph}$ give the volume of a particular cell in Cartesian and spherical sky coordinates.
The distribution $R(\alpha_i,\delta_j,z_k)$ is inferred from the input tracer catalog, and is generated by assuming that the tracers' angular, $P_{ang}(\alpha_i,\delta_j)$ and redshift,$P_z(z_k)$ probability density distributions are factorizable (\cite{demina2018computationally}):
\begin{align}
R(\alpha_i,\delta_j,z_k) = N_{tot} [ P_{ang}(\alpha_i,\delta_j) \times P_z(z_k) ] \ , 
\label{eq:factorizablility}
\end{align} 
where $N_{tot}$ is the number of tracers in the input catalog. If weights are available in the input catalogs, both $N_{wtd}$ and $R$ could be weighted. A 3-dimensional local density difference histogram $M(X_i,Y_j,Z_k)$ is then defined on the grid to represent the difference in density between $N_{wtd}$ and $R$:
\begin{align}
M(X_i,Y_j,Z_k) =  N_{wtd}(X_i,Y_j,Z_k) - R(X_i,Y_j,Z_k)\ . 
\label{eq:densitycont}
\end{align}

\subsection{The Kernel} 
\label{subsec:kernel}

To probe for BAO bubbles, we begin by creating a spherical kernel, $K(X_i,Y_j,Z_k)$, or a template to roughly model the distribution of matter expected around BAO centers, given a hypothesized BAO scale, $R_0$.
The kernel is constructed on a cube of the size of $2R_0$ in all directions, just large enough to encompass a sphere of radius $R_0$. The grid defined on this cube has the same spacing, $d_c$, as the one used to construct $M(X_i,Y_j,Z_k)$. Grid cells intersected by a sphere of radius $R_0$  centered on the center of the cube, are assigned a value of 1. All other cells are 0. 

\subsection{Convolution Step} 
\label{subsec:conv}
\begin{figure}
\includegraphics[width=\columnwidth]{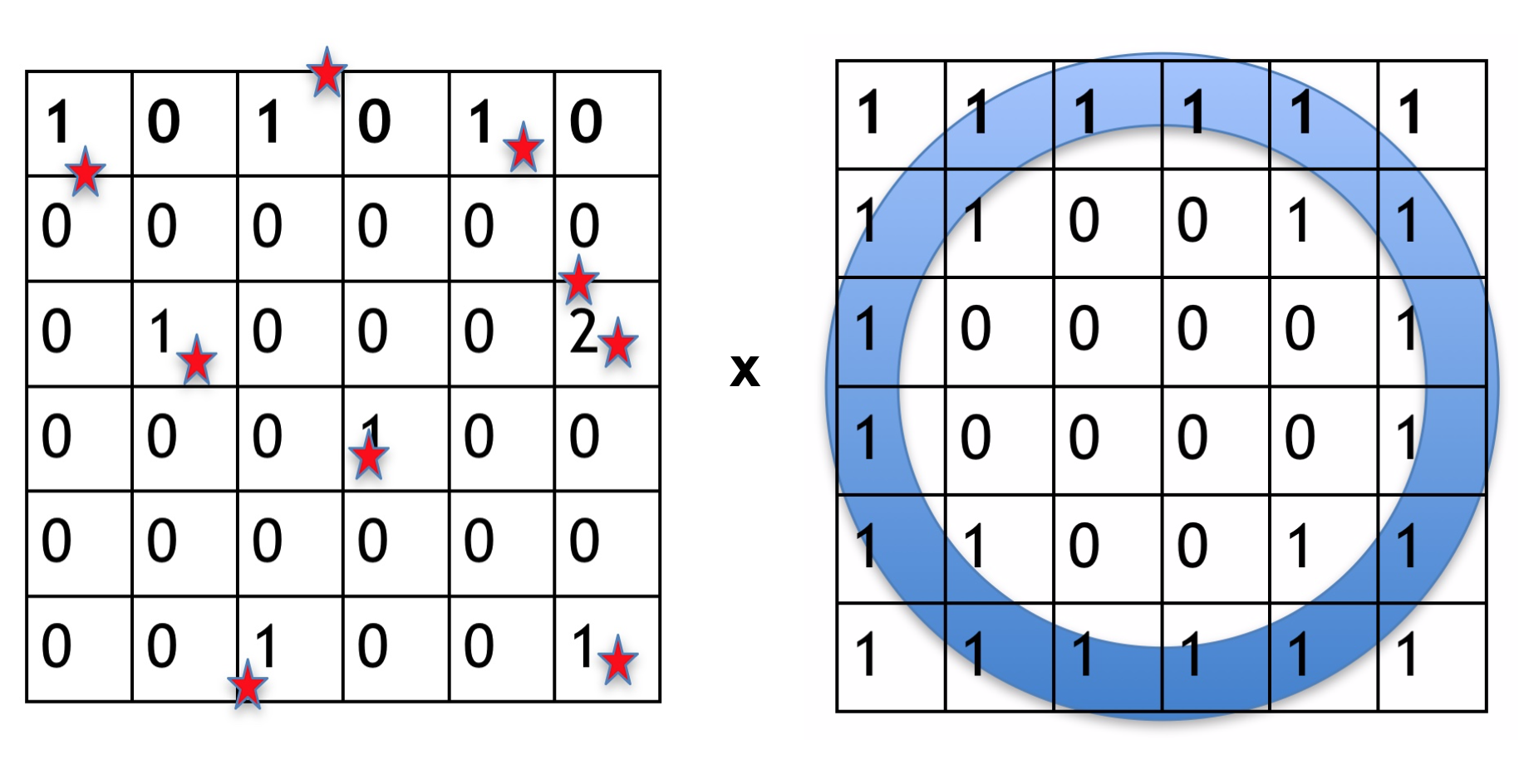}
\caption{A 2D representation of the tracer density histogram (left), being convolved with the spherical kernel (right). In this example the tensor product is 7.}
\label{fig:cf_method}
\end{figure}

At this step we construct a 3-dimensional histogram $C(X_i,Y_j,Z_k)$, with entries that quantify the likelihood that a particular point in space $(X_i,Y_j,Z_k)$ hosts a BAO center. 
The kernel $K$ is first centered on some cell $i,j,k$ and
the value of $C(X_i,Y_j,Z_k)$ is calculated as the scalar product of the kernel and the density map $M$. The kernel is then moved to a new cell, and the process is repeated until the whole surveyed volume is covered. Thus,  $C(X_i,Y_j,Z_k)$ is a result of the convolution of the matter density map $M$ with the kernel, $K$: 
\begin{align}
C(X,Y,Z) \equiv M(X,Y,Z)\ast\ast\ast K(X,Y,Z) \ , 
\label{eq:countconv}
\end{align} 
where $\ast\ast\ast$ denotes a 3-dimensional discrete convolution. 
One step in this process is visualized in Fig.~\ref{fig:cf_method}. 

If the original density histogram were the raw count of tracers, the value of $C$ would have a simple meaning: it gives the number of tracers ``voting'' for a particular cell to host a center. A larger number of ``voters'' indicates a higher likelihood of a BAO center. While $C$ does not directly correspond to ``voters'' when using the density estimate in eq.~\ref{eq:densitycont}, large values of $C$ still correspond to increased likelihood of hosting a BAO center at a particular location.

\begin{figure}
\includegraphics[width=\columnwidth]{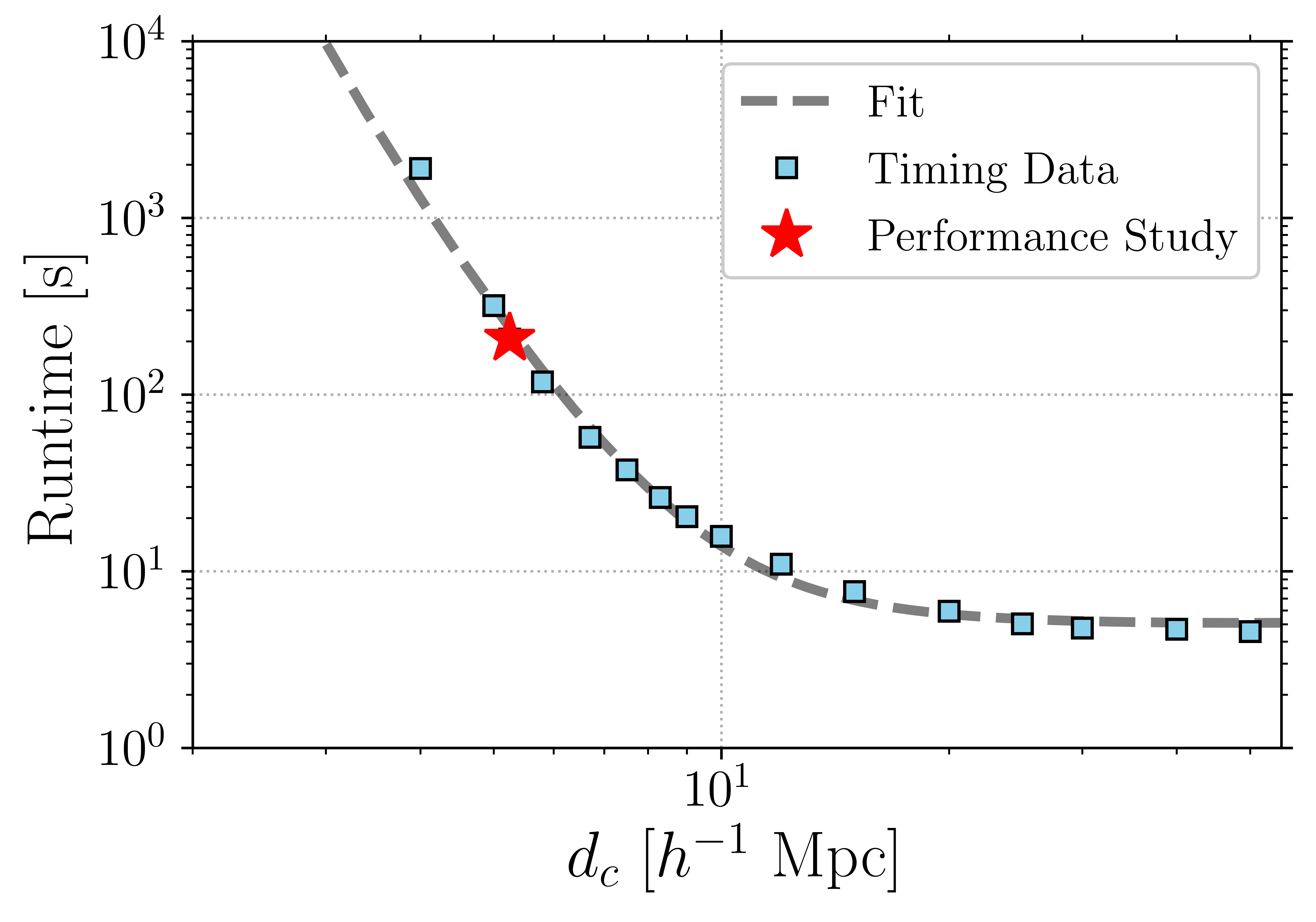}
\caption{The runtime of {\tt CenterFinder}, given as a function of the grid length, for one of the CMASS mocks used in this study (blue squares). The red star indicates the value of $d_c$ chosen in this analysis. The timing data is fit to a log power law of the form $A (d_c/d_{\mathrm{scale}})^{\alpha + \beta \log(d_c/d_{\mathrm{scale}})}+C$ (grey dashed line). In the above fit, the value of the parameter $\alpha$ is approximately $-5$. }
\label{fig:runtime}
\end{figure}

This highly vectorized convolutional algorithm, called {\tt CenterFinder}, is quite efficient; the runtime is set by the number of cells in the surveyed volume, which is in turn set by the one dimensional grid length. Shown in Fig.~\ref{fig:runtime}, it decreases with the grid length until it saturates at around 5 s. Prior to the saturation, the runtime scales as a power law, approximately $d_c^{-5}$ In our performance study (\S~\ref{sec:performance}), a grid length of $d_c = 5.25$ $h^{-1}$ Mpc yields a runtime of approximately 240 s. This study applies {\tt CenterFinder} to the northern galactic cap (NGC) catalog of the SDSS DR9 CMASS survey using a personal computer with a 2.3 GHz Intel Core i7 processor and 8 GB of memory.

\subsection{Output} 
\label{subsec:output}

Once the 3-dimensional histogram $C$ is generated, it is converted into a catalog of probable BAO centers by applying a user-specified threshold, $C_{min}$, on $C$:
\begin{align}
C(X_i,Y_j,Z_k) \geq C_{min} \ .
\label{eq:cutoff}
\end{align} 
The output catalog in {\tt astropy FITS} format provides the locations (right ascensions, declinations and redshifts) of the probable centers, and weights corresponding to the values of $C$. The redshifts are calculated using the inverse of the relation in Eq.~\ref{eq:hubbleint}. Redshifts, rather than radial distances, are included since most algorithms for probing cosmological structure are designed to be applied to redshift catalogs.
\section{Performance Study}
\label{sec:performance}

\subsection{Galaxy Survey Data, Mock and Random Catalogs}
\label{subsec:galaxies}
The  performance of {\tt CenterFinder}, i.e. its ability to locate BAO centers, is tested on the SDSS DR9 CMASS galaxy survey \citep{ahn2012ninth,padmanabhan20122} and on an ensemble of associated mock and random catalogs. We limit our analysis to galaxies in the survey’s north galactic cap. Details regarding the mock catalogs may be found in \cite{manera2013clustering}. The mock ensemble consists of 20 catalogs. The random ensemble also consists of 20 catalogs of  density equal to that of the mocks, appropriately sampled from the several large random catalogs. 

\subsection{Cosmology and the two point correlation function}
\label{subsec:cosmo2pt}

In this study we used the following values for cosmological parameters: $c = 300000$ km$/$s, $H_0 = 100h$ km$/$s$/$Mpc, $\Omega_{M} = 0.274$, $\Omega_{\Lambda} = 0.726$, and $\Omega_{k} = 0$. Before analyzing the catalogs with {\tt CenterFinder}, we investigate the clustering behavior of the DR9 galaxies and mocks with the two point correlation function (2pcf), $\xi(s)$.  We compute $\xi(s)$ using the estimator of Landy and Szalay \citep{Landy:1993yu,Hamilton:1993fp},

\begin{align}
\hat{\xi}(s) = \frac{DD(s) - 2DR(s) + RR(s)}{RR(s)} \ ,
\label{eq:2pcfdef}
\end{align}
where $DD$, $RR$, and $DR$ are the normalized distributions of the pairwise combinations of galaxies from ``data'', $D$ and ``random'', $R$ catalogs (plus cross terms) at given distances $s$ from one another. The 2pcf calculations are done using the algorithm described in \cite{demina2018computationally}.

\begin{figure}
\includegraphics[width=\columnwidth]{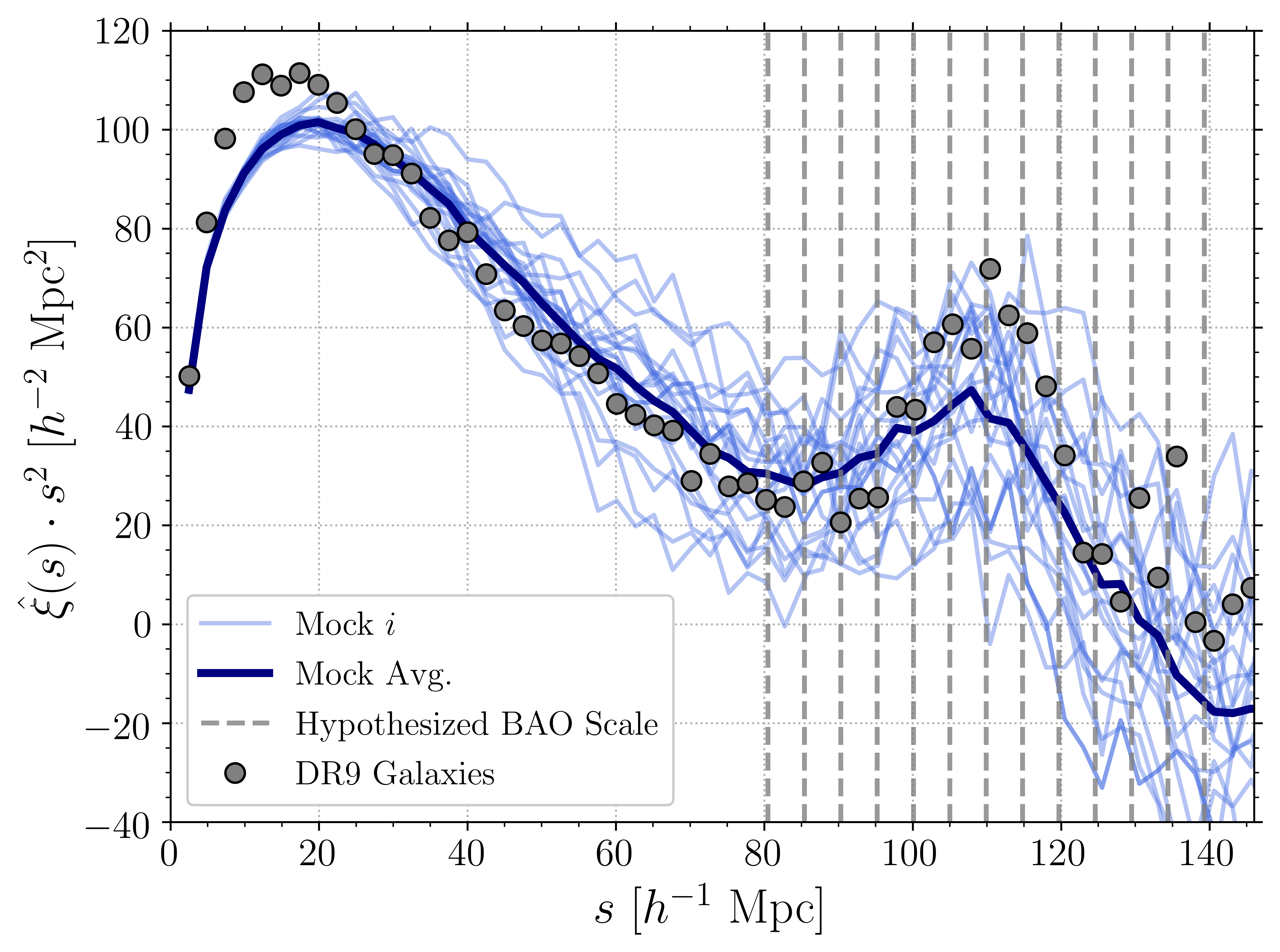}
\caption{The 2pcf, $\hat \xi (s)$, of SDSS DR9 galaxies (grey circles) and 20 mock catalogs (blue lines) described in \S~\ref{subsec:galaxies}, as a function of the comoving separation, $s$. The average of the mock ensemble is given by the solid navy line. Kernel sizes (hypothesized BAO scales) used by {\tt CenterFinder} in this analysis are shown by vertical grey dashed lines.}
\label{fig:2pt}
\end{figure}

The 2pcf calculated using bin width $\Delta s = 2.5$ $h^{-1}$Mpc is shown for data and mocks in Fig.~\ref{fig:2pt}. In both, there is evidence of clustering at small scales and a prominent BAO signal around $109$ $h^{-1}$Mpc, consistent with previous analyses on the same data sets~\citep{anderson2012clustering,ross2012clustering}. We note that both at small scales and near the BAO scale, the magnitude of $\hat \xi$ is noticeably smaller in the  average over mocks compared to the DR9 survey data. Additionally, the variance between mocks in this ensemble is considerable at separations approaching the BAO scale.
\subsection{CenterFinder Procedure}
\label{subsec:cfparams}

To identify a BAO signal using {\tt CenterFinder}, we apply it to SDSS DR9 CMASS data, mock and random catalogs. Random catalogs match both the fiducial volume and the galaxy density of the survey. 
The cosmology used is identical to the one used in the calculation of the 2pcf.

\begin{figure}
\includegraphics[width=\columnwidth]{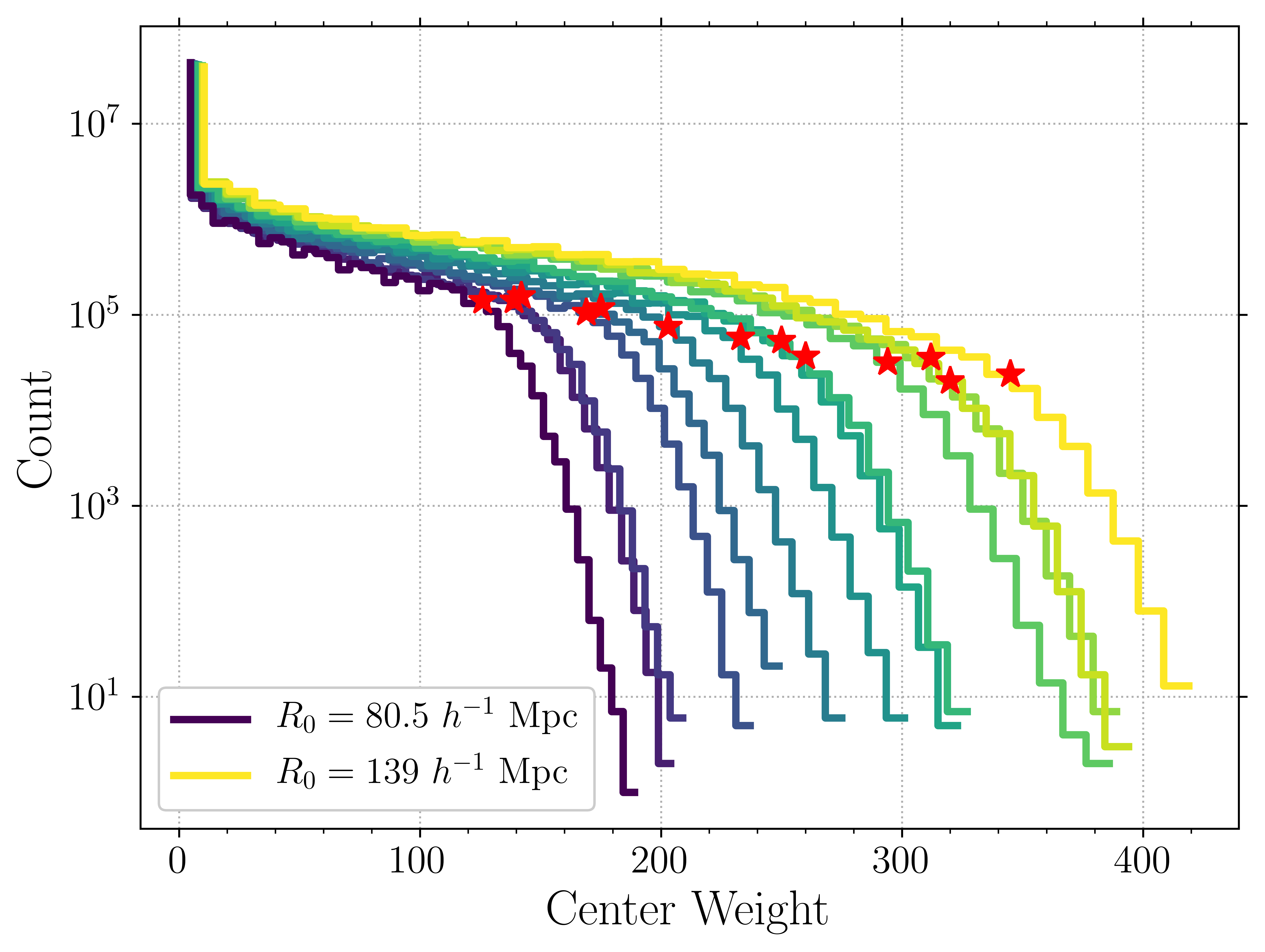}
\caption{The distribution of center weights, for kernels of different sizes, applied to a random catalog with the same fiducial volume and density as the SDSS DR9 CMASS NGC survey. The dark blue curve corresponds to $R_0 = 80.5$ $h^{-1}$ Mpc and yellow corresponds to $R_0 = 139$ $h^{-1}$ Mpc. Intermediate colors correspond to the intermediate kernel sizes used in this analysis. The red stars indicate the threshold values, $C_{min}$ used to create catalogs of probable centers at each kernel size.}
\label{fig:random_wts}
\end{figure}

We generate a density map $M$ using the method described in \S~\ref{subsec:density}, with a grid spacing, $d_c=5.25$ $h^{-1}$Mpc.  The radius of the kernel, $R_0$, is varied from $80.5-139$ $h^{-1}$Mpc in $4.9$ Mpc steps. For each value of $R_0$, the kernel is convolved with the density map $M$ to generate a map of weights, $C({R_0})$, where larger values indicate a higher probability of BAO centers being present at that location in space.  The distribution of the values of $C(R_0)$ from a random catalog is shown in Fig.~\ref{fig:random_wts} for different kernel sizes.

 These distributions have three prominent features. First, the vast majority of cells in the grid are empty or nearly empty, and their convolution with the kernel yields small values of $C$, as shown by the sharp spike at low weights. It is followed by gradual falling off, which extends to higher values for larger kernels. Finally, there is a sharp drop in counts as the weights increase.

\begin{figure}
\includegraphics[width=\columnwidth]{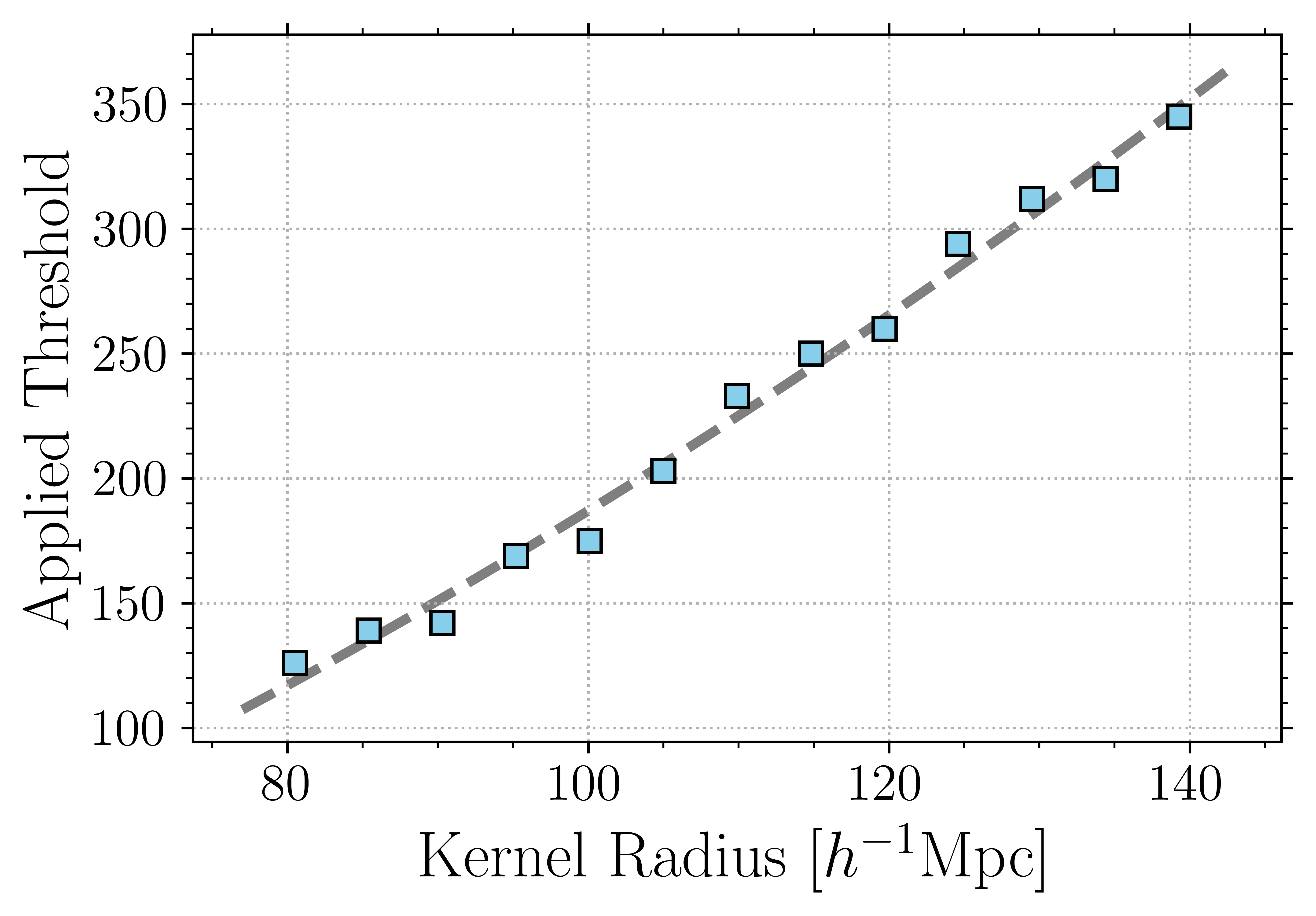}
\caption{The threshold values, $C_{min}$, applied to our center catalogs, as a function of the kernel size (blue squares). The points are fit to a second order polynomial (dashed grey line).}
\label{fig:cuts}
\end{figure}

 To reduce the size of the output catalog we keep only the cells with values larger than a certain threshold, $C_{min}$, which is chosen to keep the count of centers approximately constant for different values of $R_0$. Since the number of counts on a spherical shell is proportional to the radius squared, we expect $C_{min}$ to be approximately proportional to $R_0^2$. We choose values of $C_{min}$ (shown by red stars in Fig.~\ref{fig:random_wts}), which corresponds to the beginning of a sharp drop off. We apply a threshold selection at $C_{min} = 150$ for the smallest kernel size and  at $C_{min} = 300$ for the largest. The dependence of the chosen thresholds on the kernel size shown in Fig.~\ref{fig:cuts}. As expected, this dependence is well fit by the second order polynomial.

Finally, we generate FITS catalogs of the probable centers based on 20 mock, 20 random, and DR9 data catalogs for several values of the kernel size.

\subsection{An Illustration of CenterFinder}
\label{subsec:illust}

\begin{figure}
\includegraphics[width=\columnwidth]{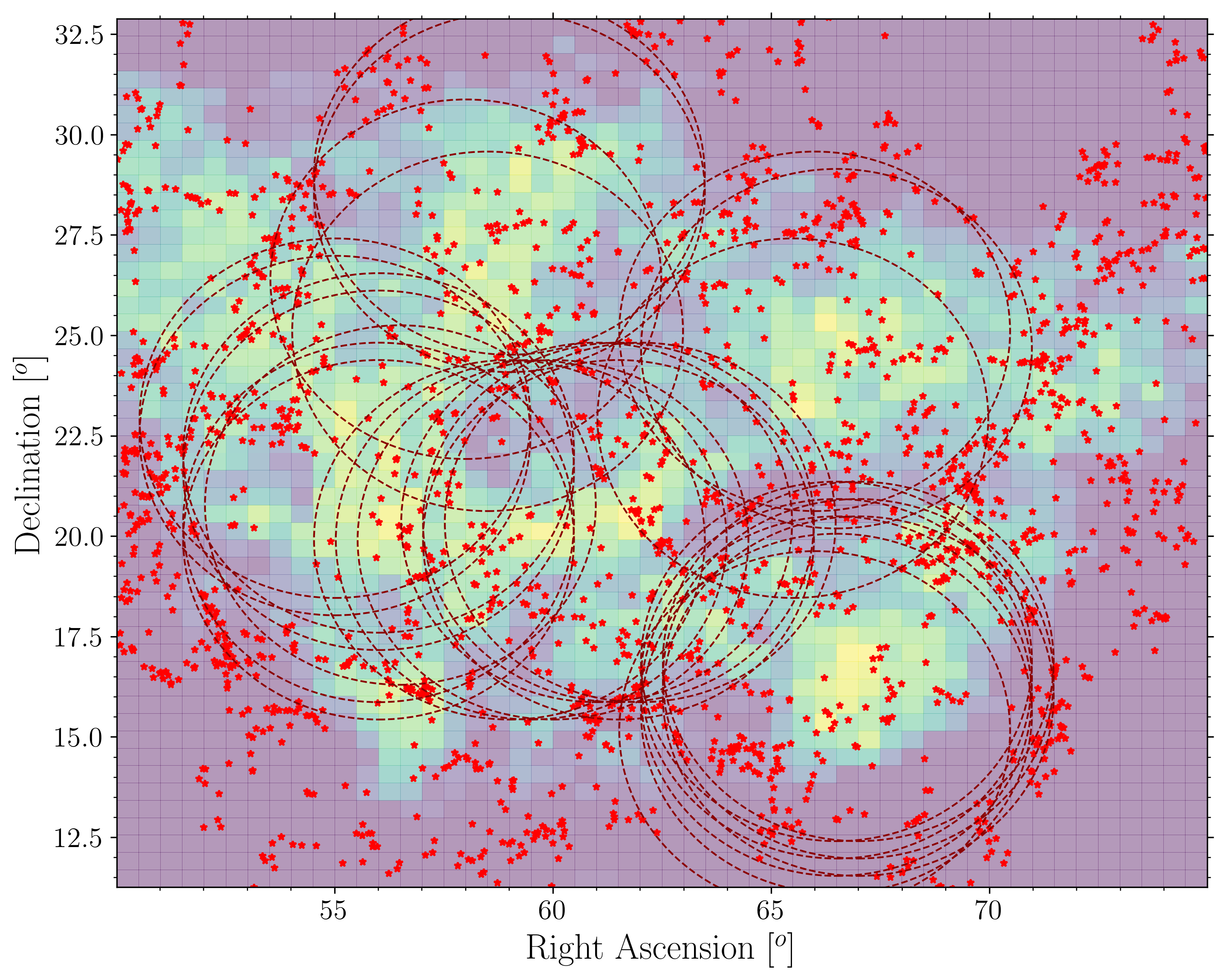}
\caption{The angular distribution of the probable centers for the redshift range $0.51<z<0.52$ (color scale), generated from the $R_0=109.9$ $h^{-1}$ Mpc kernel  on a portion of the SDSS DR9 data. Galaxies locations are shown as red stars. The dashed circles reflect the angles subtended by the BAO spheres at $z=0.515$. BAO circles are only shown around the most probable center locations.}
\label{fig:illust}
\end{figure}

Using a sample of galaxies from the SDSS DR9 survey as an example, we illustrate the output of {\tt CenterFinder}. In Fig.~\ref{fig:illust} we show a slice in the red-shift region from $z=0.51$ to $z=0.52$ from the catalog of probable centers generated using the kernel size of $R_0=109.9$ $h^{-1}$ Mpc.  The color of the background represents the weights $C$, which characterize the probability that a particular point is a location of a BAO center with more intense yellow being the most probable. Red points show the location of the galaxies from the SDSS DR9 catalog from the same fiducial volume. Dashed  lines are circles of $109.9$ $h^{-1}$ Mpc radius around the probable BAO centers.  Even in this 2-dimensional slice, we observe significant overlap between the circles denoting the BAO shells, and the galaxies from the survey. Since we chose a fairly low value for the threshold selection of BAO centers there are multiple circles around the same location with high probability to host the BAO center. It is also noticeable that these locations typically host several galaxies, supporting the hypothesis that dark matter enriched BAO centers are likely to be seeds for galaxy formation. 

\subsection{Probable Center to Galaxy Cross Correlations}
\label{subsec:xcorr}

Since potential centers correspond to primordial over-dense regions, we expect the galaxies to have been preferentially formed in these regions of space. We test this hypothesis by cross correlating the potential BAO centers, $D_C$, with  galaxies, $D_G$ drawn from data or mock catalogs. For each catalog of centers, generated using a kernel of size $R_0$, we calculate a cross correlation function, $\hat \xi_{GC} (s,R_0)$:

\begin{align}
\hat{\xi}_{GC}(s,R_0) =  \frac{D_GD_C - D_GR_C- D_CR_G + R_GR_C}{R_GR_C} \ ,
\label{eq:xcorr}
\end{align}
where $R_C$ and $R_G$ are random catalogs corresponding to the centers and galaxy catalogs respectively. All pairwise distributions are functions of the distance between a center and a galaxy, $s$ and the hypothesized BAO scale, $R_0$. 

$R_G$ corresponds to the random galaxies used in the evaluation of the 2pcf (\S~\ref{subsec:cosmo2pt}). $R_C$ is a catalog of $N_{CR}$ randomly distributed centers. The ratio of $N_{CR}$ to the number of found centers is chosen to be similar to the ratio between the number of galaxies in data $D$ and the size of the random catalog $R$. $R_C$ is generated by applying {\tt CenterFinder} to several random galaxy catalogs. The angular and redshift distributions of $R_C$ match those of the probable centers, $D_C$ as detailed in in App.~\ref{app:dist}.

\begin{figure}
\includegraphics[width=\columnwidth]{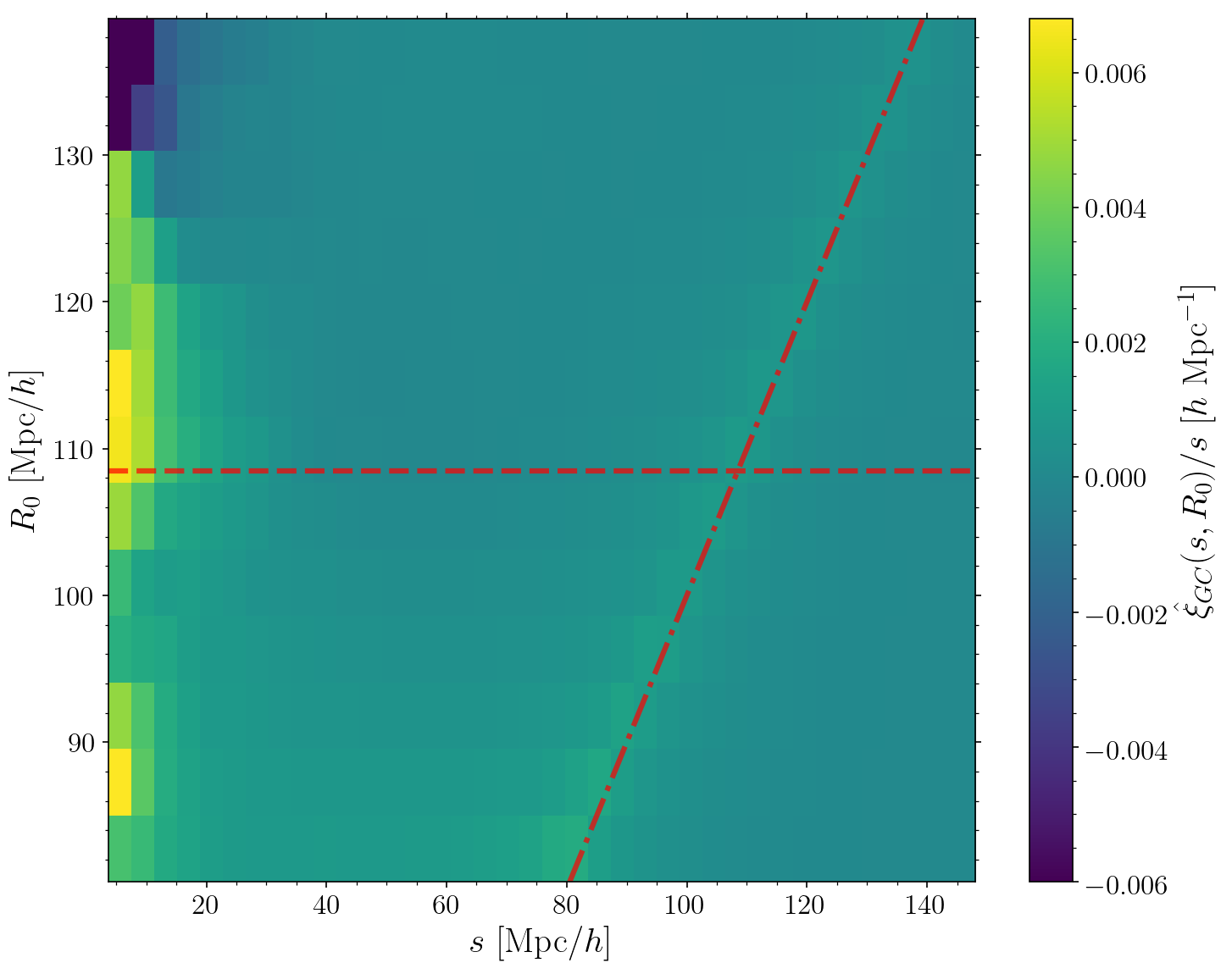}
\caption{The cross correlation of probable center locations with SDSS DR9 galaxies as a function of the objects' separation, $s$, and the hypothesized BAO scale of {\tt CenterFinder} used to generate the center catalog, $R_0$. The cross correlation is adjusted for clarity (divided by $s$). The dashed line (red) marks an estimate of the BAO scale observed in the 2pcf. The dot-dash line (red) marks $s=R_0$.}
\label{fig:dataxcorr}
\end{figure}

\begin{figure}
\includegraphics[width=\columnwidth]{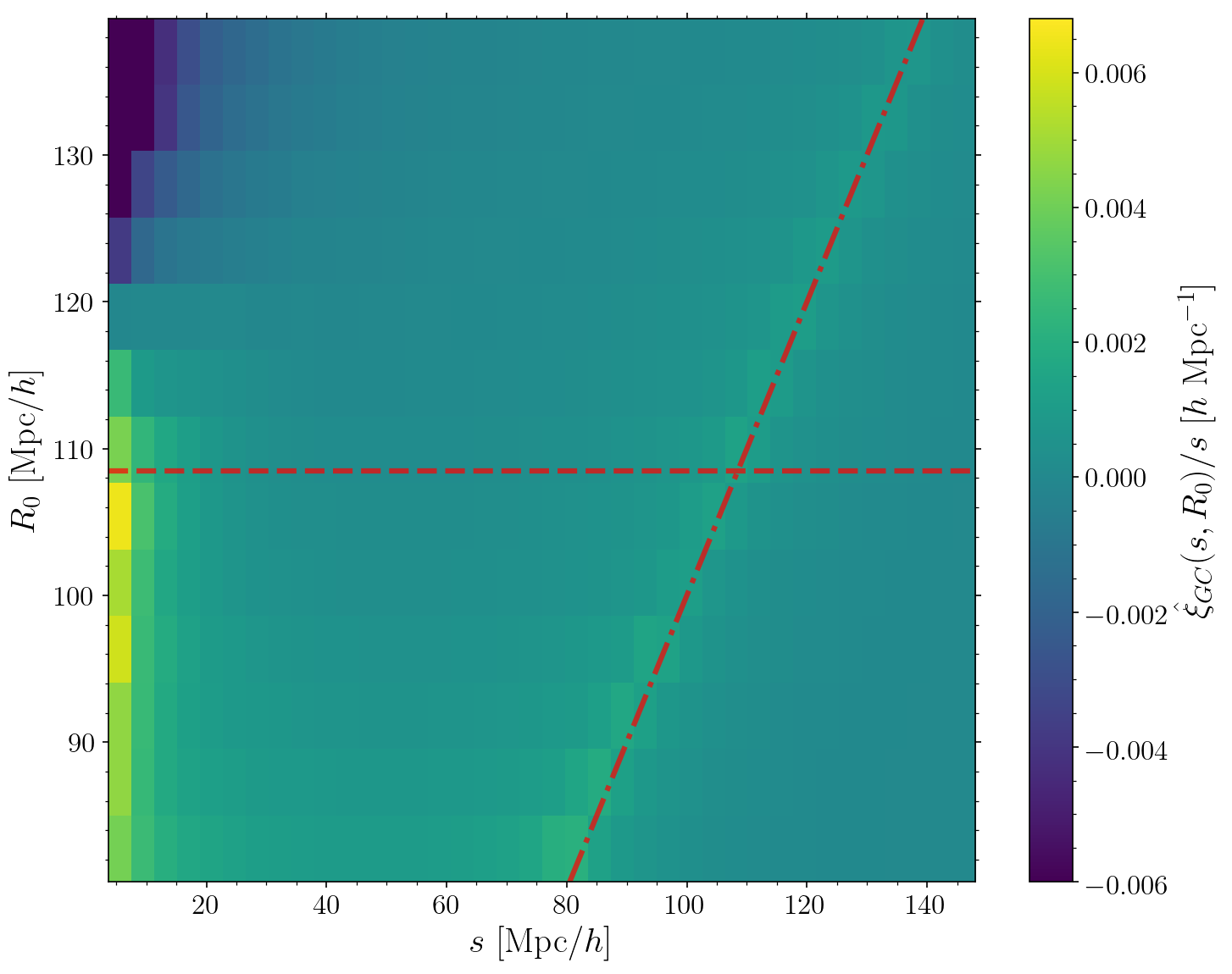}
\caption{
Same as in Fig.~\ref{fig:dataxcorr} for mock galaxies.}
\label{fig:mockxcorr}
\end{figure}

\begin{figure}
\includegraphics[width=\columnwidth]{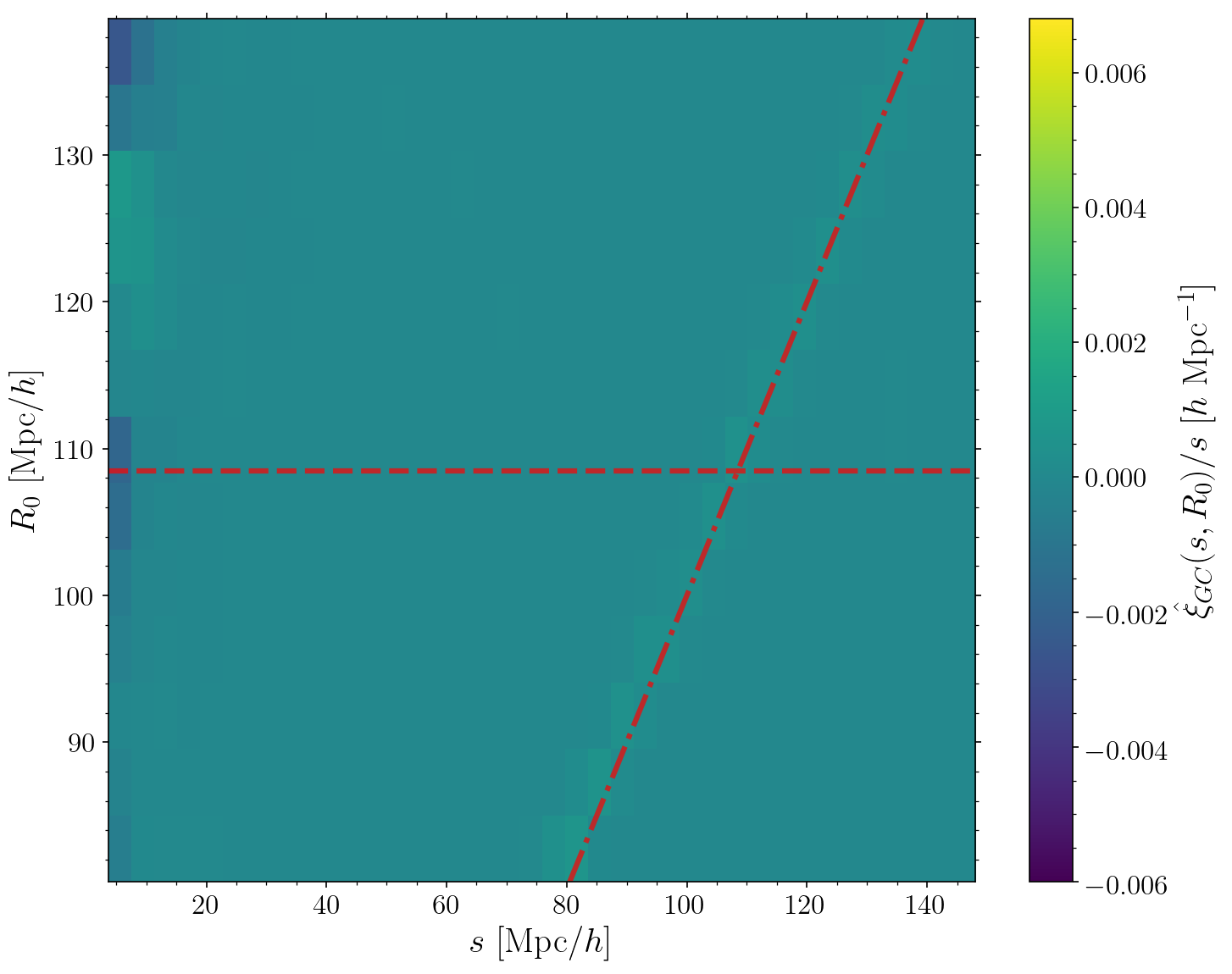}
\caption{
Same as in Fig.~\ref{fig:dataxcorr} for random galaxies.}
\label{fig:randomxcorr}
\end{figure}

$\hat{\xi}_{GC}(s,R_0)$ is evaluated using {\tt nbodykit}, which is described in \cite{hand2018nbodykit}.
Figs.~\ref{fig:dataxcorr}, \&~\ref{fig:mockxcorr} and ~\ref{fig:randomxcorr} show the result of the center-galaxy cross correlation in data, mock and random catalogs respectively. The distributions are divided into $s$ bins of width $\Delta s = 3.8$ $h^{-1}$Mpc and $R_0$ bins reflecting the sampling when applying {\tt CenterFinder}. 

Two clustering features are present in Figs.~\ref{fig:dataxcorr} \&~\ref{fig:mockxcorr}. First, a region of higher values of $\hat \xi_{GC}$ which follows $s = R_0$ is observed in both plots, with roughly equal magnitude across mock  and SDSS DR9 data. This feature is an artifact of the {\tt CenterFinder}.  The algorithm is designed to find centers of  densely populated spherical shells of a given radius. The observed excess is simply the correlation between the galaxies in these shells with the corresponding found centers. This feature confirms the most basic function of the algorithm, but is not physically informative. 

The second feature in both  mock and  DR9 survey data is the observed   increase in $\hat \xi_{GC}$ at low $s$, which is enhanced  when the hypothized BAO radius is near the true BAO scale (horizontal dashed line) \citep{anderson2012clustering,ross2012clustering}, and diminished at larger and smaller radii. This behavior confirms our original hypothesis that galaxies cluster around the potential BAO centers. 
The first (diagonal) feature corresponding to the kernel radius is observed in random catalogs (Fig.~\ref{fig:randomxcorr}), but there is no excess at the small scales. This qualitatively confirms that the signal at low $s$ seen in the mocks and in the data is due to large scale structures and is not a random fluctuation.

\begin{figure}
\includegraphics[width=\columnwidth]{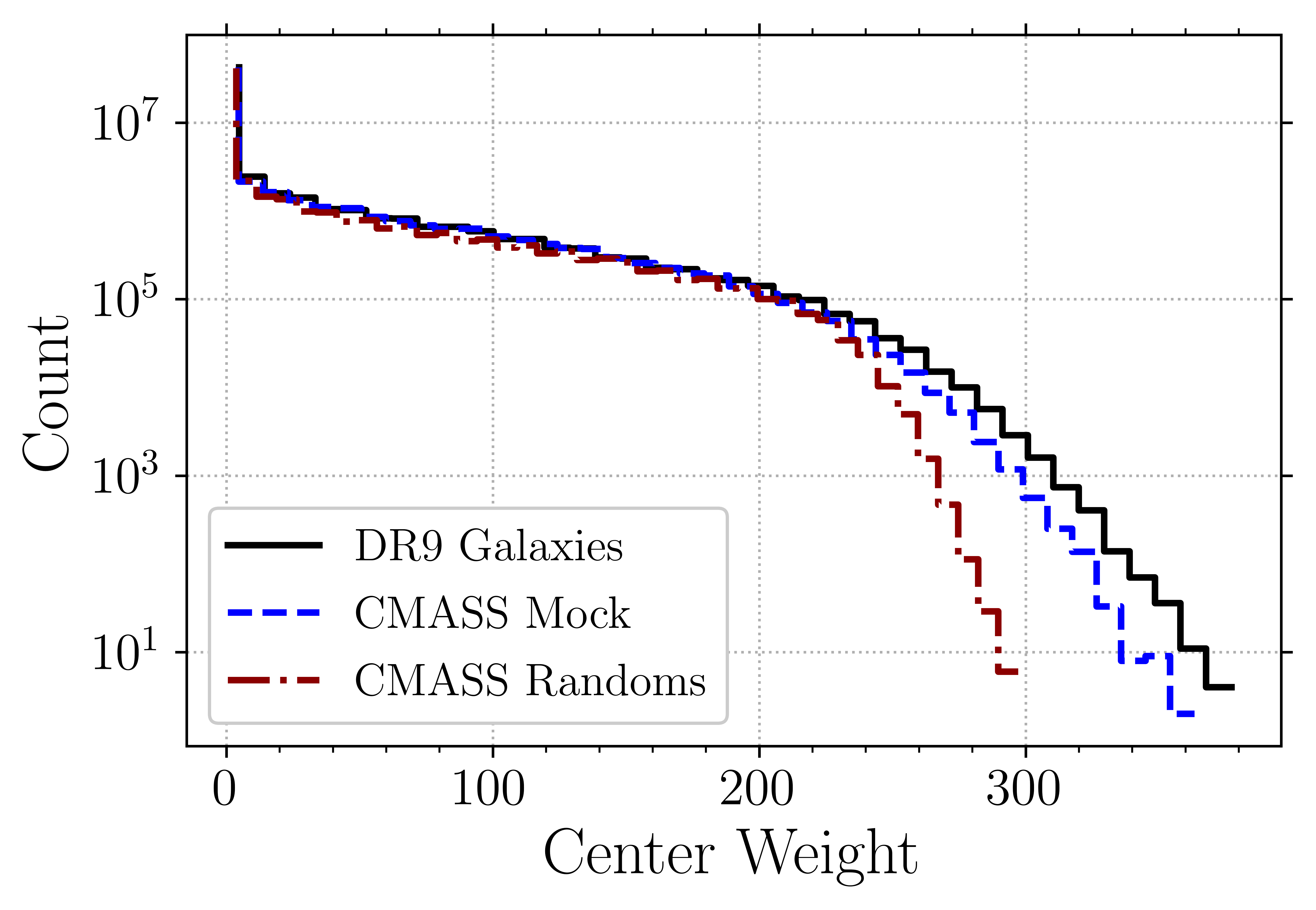}
\caption{The distribution of center weights, for a kernel with a radius reflecting the approximate BAO scale, extracted from the 2pcf, $R_0 = 109.9$ $h^{-1}$ for SDSS DR9 galaxies (black solid line), one of the CMASS mocks (blue dashed line), and one of the CMASS randoms (red dot dash line).}
\label{fig:wts}
\end{figure}

We note that the BAO signal (increase in $\hat \xi_{GC}$ at low $s$) is larger in magnitude for SDSS DR9 data (Fig~\ref{fig:dataxcorr}) compared to the average over the mock catalogs (Fig~\ref{fig:mockxcorr}). This is consistent with the fact that the mock catalogs showed on average less overall galaxy clustering in the 2pcf than the SDSS DR9 data (Fig.\ref{fig:2pt}). This is also observed in the distribution of probable center weights, $C$, shown in Fig.~\ref{fig:wts}. The distribution in Fig.~\ref{fig:wts} are shown prior to applying a threshold, $C_{min}$.

The SDSS DR9 galaxies show higher counts of probable centers with larger weights, or values of $C$. This suggests a higher degree of galaxy density and clustering on the BAO shells. This observation is consistent with the increased clustering behavior seen in the 2pcf. Furthermore, the distributions of probable center weights in Fig.~\ref{fig:wts} show that the SDSS DR9 galaxies and CMASS mock significantly deviate from the distribution extracted from randoms at large weights. This indicates the presence of BAO shells imprinted in the mock and data galaxy distributions.

\subsection{Quantifying the BAO Signal}
\label{subsec:quantBAO}

\begin{figure}
\includegraphics[width=\columnwidth]{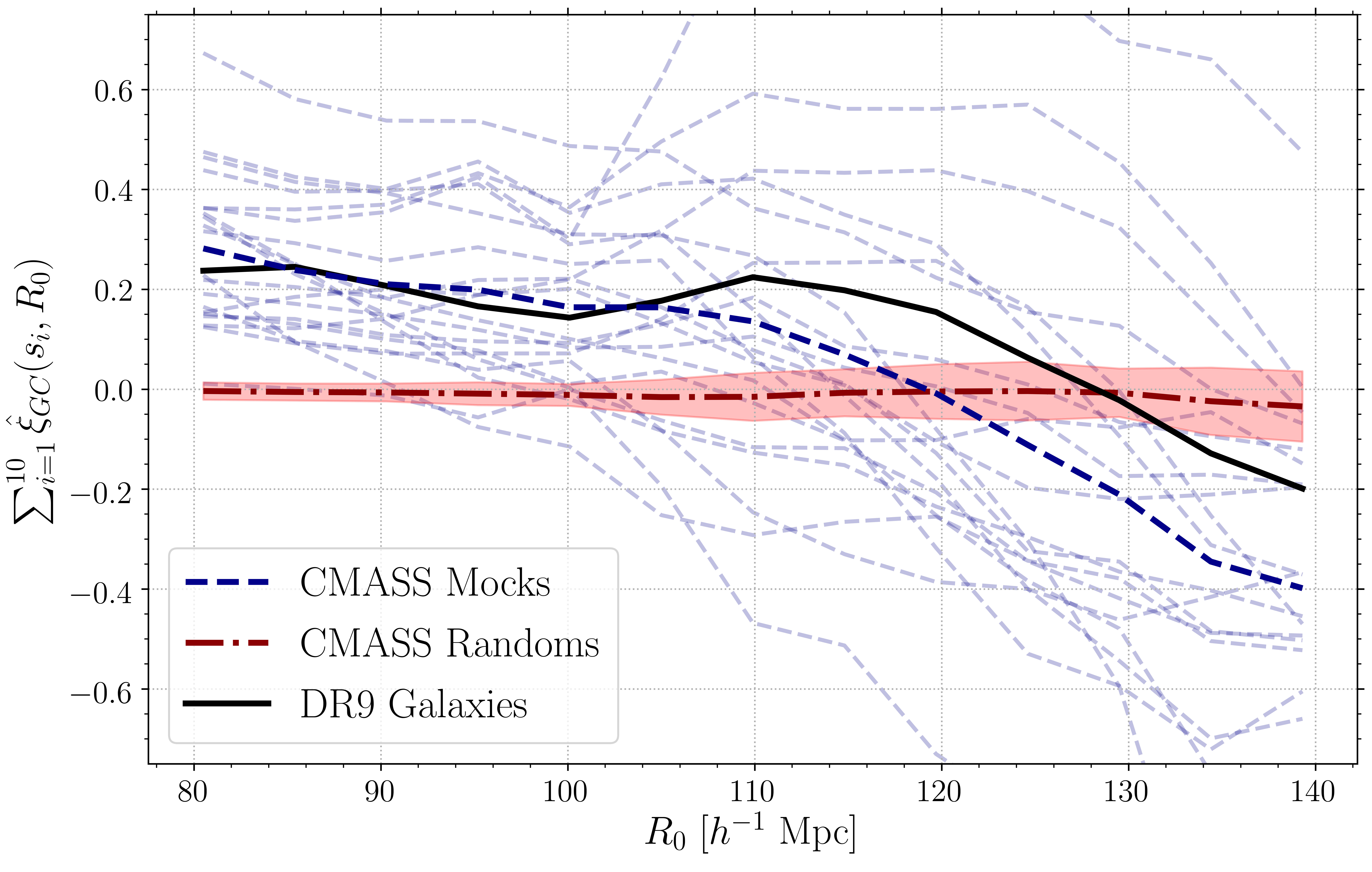}
\caption{ Signal strength as a function of kernel size. Each mock is shown individually (blue), as is DR9 data (black), while randoms (red) are represented by a mean of all values at each kernel size, with a shaded region given by the standard deviations.}
\label{fig:sigvk}
\end{figure}

\begin{figure}
\includegraphics[width=\columnwidth]{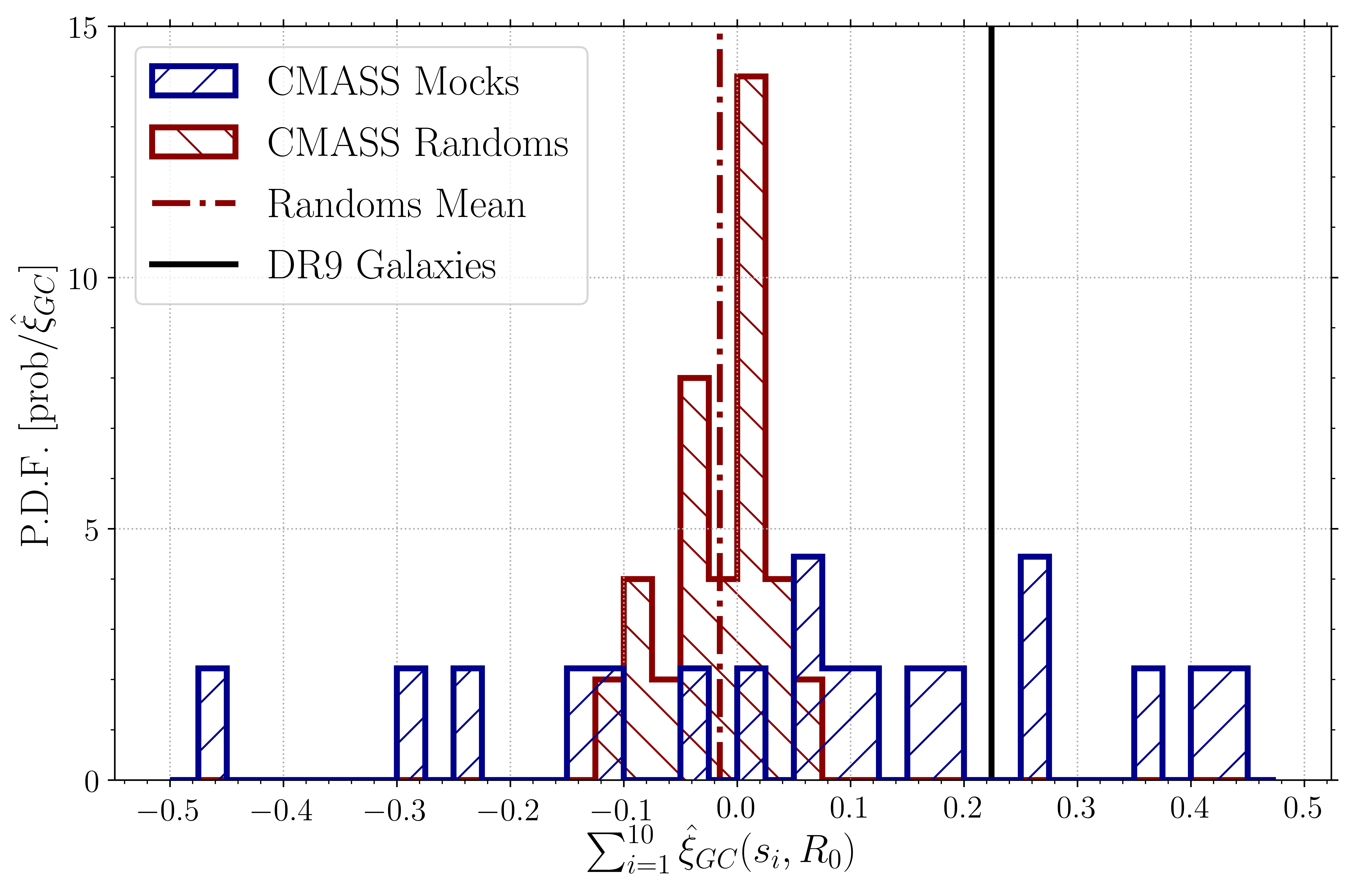}
\caption{A normalized histogram of the signal strengths at $R_0 = 109.9 \ h^{-1}$ Mpc for the CMASS Randoms (red) and CMASS Mocks (blue). The SDSS DR9 data (black) is shown as a vertical line at $0.224$, $5.0$ standard deviations away from the random mean at $-0.015$. }
\label{fig:sighist}
\end{figure}

To quantify the strength of the presumed BAO signal we sum the values of $\hat \xi_{GC}$ in the first 10 $s$-bins (from $s=3.72$  to $s=42.2$ $h^{-1}$ Mpc) for every $R_0$ value. The results are shown in Fig.~\ref{fig:sigvk} for the SDSS DR9 data, mock, and random catalogs.
The distribution over signal strengths for mocks and randoms, as well as the value for SDSS DR9 data at $R_0 = 109.9$ $h^{-1}$ Mpc is shown in Fig.~\ref{fig:sighist}.
There is a  large spread in signal strength for the mocks at every kernel size which is in agreement with the behavior observed in the two-point statistics (Fig.~\ref{fig:2pt}). The signal strength derived from random catalogs on the other hand is centered around 0 and shows a much smaller spread.  At the kernel size corresponding to the BAO scale, $R_0 = 109.9$ $h^{-1}$ Mpc, we find that the SDSS DR9 data differs from the mean of the CMASS randoms by $5.0$ standard deviations excluding the probability of a random fluctuation at more than 99\% CL.

\section{Discussion}
\label{subsec:discussion}
An approach similar to this work was suggested in \cite{arnalte2012wavelet}. In this study the kernel shape is more complex, emulating a spherical wavelet template. Our algorithm also contains several different optional kernel shapes, including a similarly shaped wavelet. 

Another difference is that {\tt CenterFinder} scans over the entire surveyed volume identifying BAO centers that may or may not be associated with other galaxies, while the algorithm described in \cite{arnalte2012wavelet} uses Luminous Red Galaxies as seeds to search for spherical shells. Hence the output of {\tt CenterFinder} can be cross correlated with other matter tracers, such as Lyman-$\alpha$ forest, and weak lensing dark matter maps. 
 
\section{Conclusions}
\label{sec:conclusion}
In this paper we present 
the algorithm {\tt CenterFinder} designed to locate centers of spherical shells generated by Baryon Acoustic Oscillations. So far the BAO signature was observed as a statistical feature in the CMB power spectrum, and in the two point correlation function of galaxy distributions.   This algorithm is computationally efficient and can be applied to a variety of tracer catalogs. A performance study using  SDSS DR9 survey and mock catalogs yielded a novel method to detect the BAO scale, and to generate catalogs of probable BAO center locations to study in future analyses. 
Using these catalogs of centers, cross correlations between them and other tracers of the cosmic web such as clusters, voids, Lyman-$\alpha$ forest, and weak lensing maps may be studied. 

\begin{acknowledgements}
The authors would like to thank S. BenZvi, K. Douglass and S. Gontcho A Gontcho for useful discussions and insightful questions. RD. thanks D. Bianchi, L. Samushia and Z. Slepian for their interest and helpful comments. The authors acknowledge support from the Department of Energy under the grant DE-SC0008475.0. Funding for SDSS-III has been provided by the Alfred P. Sloan Foundation, the Participating Institutions, the National Science Foundation, and the U.S. Department of Energy Office of Science. The SDSS-III web site is \url{http://www.sdss3.org/}.

SDSS-III is managed by the Astrophysical Research Consortium for the Participating Institutions of the SDSS-III Collaboration including the University of Arizona, the Brazilian Participation Group, Brookhaven National Laboratory, Carnegie Mellon University, University of Florida, the French Participation Group, the German Participation Group, Harvard University, the Instituto de Astrofisica de Canarias, the Michigan State/Notre Dame/JINA Participation Group, Johns Hopkins University, Lawrence Berkeley National Laboratory, Max Planck Institute for Astrophysics, Max Planck Institute for Extraterrestrial Physics, New Mexico State University, New York University, Ohio State University, Pennsylvania State University, University of Portsmouth, Princeton University, the Spanish Participation Group, University of Tokyo, University of Utah, Vanderbilt University, University of Virginia, University of Washington, and Yale University.
\end{acknowledgements}

\bibliographystyle{aa}
\bibliography{cf_bib}

\appendix
\section{Galaxy, Random, \& Center Distributions}
\label{app:dist}

\begin{figure}
\includegraphics[width=\columnwidth]{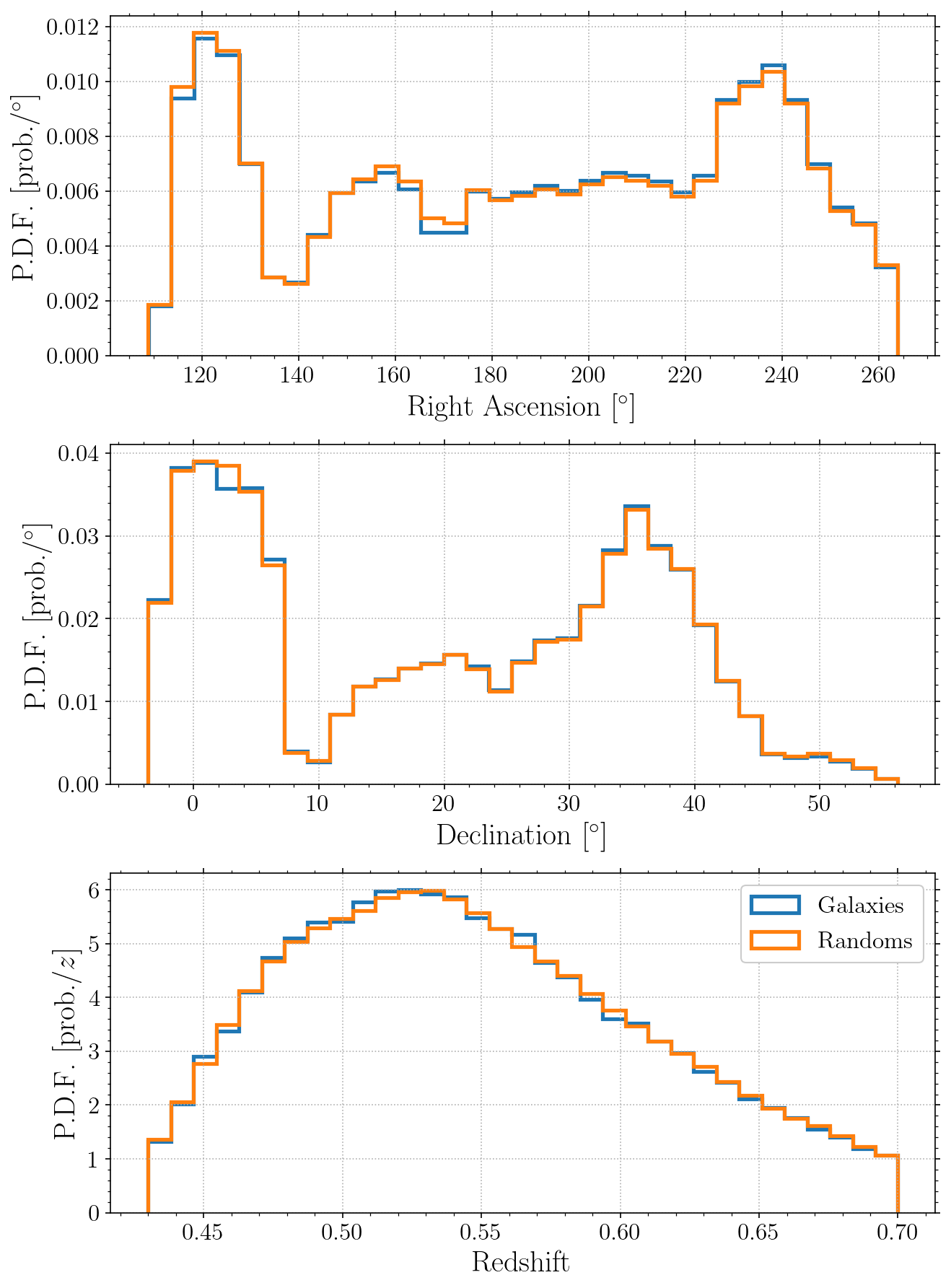}
\caption{For the SDSS DR9 CMASS galaxies (blue) and random galaxies (orange), we plot the distribution over the right ascension (top), declination (middle), and redshift (bottom).}
\label{fig:distsgal}
\end{figure}

\begin{figure}
\includegraphics[width=\columnwidth]{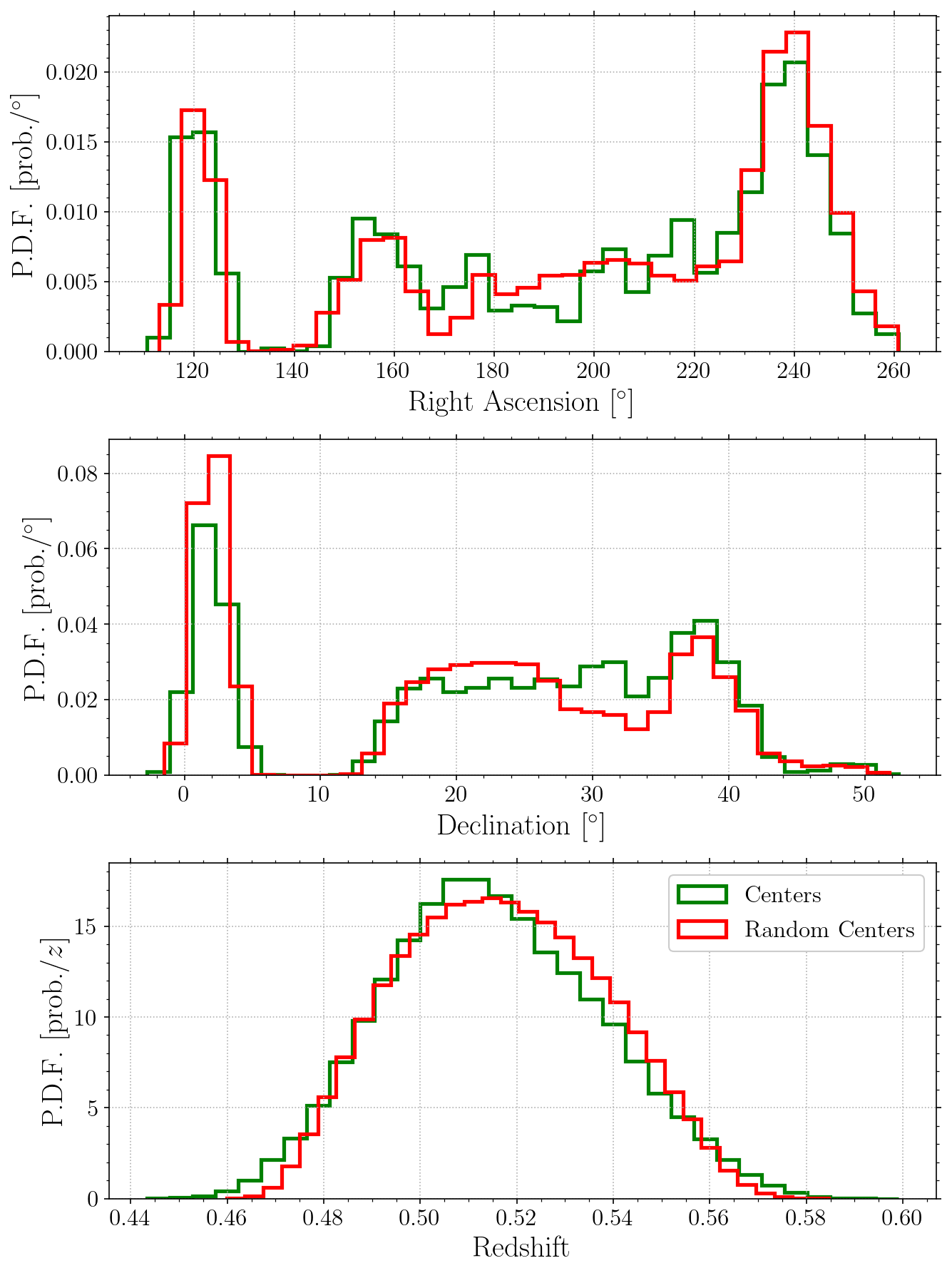}
\caption{For the probable centers at $R_0 = 109.9$ $h^{-1}$ Mpc, generated from both data (green) and randoms (red), we plot the distribution over the right ascension (top), declination (middle), and redshift (bottom).}
\label{fig:distscen}
\end{figure}

In \S~\ref{subsec:xcorr}, we cross correlate objects from two catalogs, galaxies and probable centers. In addition, they are compared to random distributions, $R_G$ and $R_G$. In Figs.~\ref{fig:distsgal} \& \ref{fig:distscen}, we show the distributions of these tracers over  their coordinates. 

The probable centers are clearly biased towards regions of the survey with high galaxy density. While the method outlined in \S~\ref{subsec:density} corrects for some of this, Figs.~\ref{fig:distsgal} \& \ref{fig:distscen} demonstrate that the effect is not entirely removed. The difference in the distribution over the redshift especially justifies use of an additional random catalog, $R_C$, in the evaluation of the galaxy to probable center cross correlation.


\end{document}